\documentclass[aps,pre,reprint,10pt,superscriptaddress,showpacs]{revtex4-1}
\usepackage{amsfonts} 
\usepackage{amsmath}
\usepackage{amssymb}
\usepackage{graphicx}
\usepackage{subfigure}
\usepackage{color}
\newcommand*\xbar[1]{%
  \hbox{%
    \vbox{%
      \hrule height 0.5pt 
      \kern0.5ex
      \hbox{%
        \kern-0.1em
        \ensuremath{#1}%
        \kern-0.1em
      }%
    }%
  }%
} 
\begin{document}
\title{Optical surface plasmons at a metal-crystal interface with the Drude-Lorentz model for material permittivity}
\author{A. P. Misra}
\email{apmisra@visva-bharati.ac.in; apmisra@gmail.com}
\affiliation{Department of Mathematics, Siksha Bhavana, Visva-Bharati (A Central University), Santiniketan-731 235,  India}
\author{M. Shahmansouri}
\email{mshmansouri@gmail.com}
\affiliation{Department of Physics, Faculty of Science, Arak University, Arak, P.O. Box 38156-8-8349, Iran}
\author{N. Khoddam}
\email{nkhoddam2002@gmail.com}
\affiliation{Department of Physics, Faculty of Science, Arak University, Arak, P.O. Box 38156-8-8349, Iran}
\begin{abstract}
The theory of surface electromagnetic waves (SEMWs) propagating at optical frequencies   along  the interface of an isotropic  metallic plasma [e.g., gold (Au)] and a uniaxial crystal [e.g., Rutile (TiO$_2$)] is revisited with the Drude-Lorentz (DL) model for the complex dielectric material permittivity $(\epsilon_p)$. The latter   accounts for the contributions of both the \textit{intraband} transitions of the free electrons and the multiple \textit{interband} transitions of the bound electrons in metals.  The    propagation characteristics of the wave vectors and wave frequency of SEMWs, the hybridization factors, i.e., the amplitude ratios between the transverse-electric (TE) and transverse-magnetic (TM) modes in the isotropic metal, and between the ordinary and extraordinary  modes in the uniaxial substrate   are studied numerically. It is found that the results are significantly modified from those with the Drude model for $\epsilon_p$,  especially in the short-wavelength spectra ($\lambda\lesssim500$ nm) and with a small deviation of the orientation of the optical axis. The excitation of such SEMWs can have novel applications in transportation of EM signals in a specified direction at optical frequencies ($\sim $ PHz).  
\end{abstract}
\maketitle
\section{Introduction} \label{sec-intro}
The interaction between electromagnetic (EM) waves and isotropic metals is  ascertained  by the collective movements of free electrons (in the long-wavelength spectra $>500$ nm) as well as multiple interband transitions of bound electrons (in the short-wavelength spectra $\lesssim500$ nm)  in metals. The optical and transport properties of the latter are usually described by the complex  permittivity function  $(\epsilon_p)$ of the wave frequency and wave vector. The simple  models for $\epsilon_p$ are, e.g., the Drude model \cite{drude1900} which account  for only the  \textit{intraband} transition of free electrons in metals.  
 The Drude's model is based on the classical mechanical theory of free electrons in metals with immobile positive ions. The Drude's oscillator is essentially an extension of the single Lorentz oscillator with no restoring force and no resonance frequency.  The model explains the transport properties of conduction electrons in metals  due to \textit{intraband} transitions, i.e., transitions between levels within a conduction or valence band in a quantum-mechanical interpretation. However, the detailed mechanism of this \textit{intraband} electronic conduction can be found from the Drude theory \cite{drude1900}.  
However, for a practical metal, in addition to this \textit{intraband} transition, there are usually multiple \textit{interband} transitions of the bound electrons.  Physically, when   electrons in bound states are excited by the high-energy photons   they can jump up from the lower energy band (below the Fermi level) to the conduction bands. This process   is known as an \textit{interband} transition of electrons   in which no energy state is allowed in between the filled valence band and an empty conduction band.  Such interband transitions frequently take   place in the light-matter interactions at the visible or infrared wavelengths.   In the short-wavelength spectra ($\lambda<500$ nm), the dielectric function can no longer be accurately described by the Drude   model  but can be well described  by the Drude-Lorentz (DL) model \cite{rakic1998} which holds all over the spectrum (i.e., from visible to near infrared wavelengths).
\par
Surface plasmon polaritons (SPPs) are typical electromagnetic (EM) waves that propagate along a metal-dielectric interface and whose  amplitudes decay exponentially  away from the interface. Because of their tighter spatial confinement and  higher local field intensity,  as well as  high sensitivity to the permittivity function, SPPs have been used in various applications including sensing \cite{homola1999,chung2011}, imaging \cite{zhang2008}, nano-photon detectors \cite{tang2008}, enhanced second harmonic generation \cite{chen1981}, surface enhanced Raman scattering \cite{metiu1982}, and many more. Such SPPs propagate not only at  the interface of an isotropic metal and an isotropic dielectric \cite{wang2016}, but also at the interface between  an isotropic metal and an anisotropic dielectric \cite{li2008,moradi2018}. Examples include the Dyakonov surface waves (DSWs) \cite{dyakonov1988,averkiev1990} and Dyakonov surface plasmons (DSPs) \cite{jacob2008} which have  properties of both the Dyakonov surface waves and the SPPs. The DSWs have  some unique characteristics, e.g., they are weakly localized and they propagate at the interface of two media at least one of which is anisotropic and the real parts of the permittivity functions are of opposite sign. Also, they are  hybridized due to polarization of both the transverse electric (TE) and the transverse magnetic (TM) fields, and they are highly directional, i.e., they can exist only under certain conditions and in specific regimes \cite{takayama2017}.  Here, by TE (TM) fields we mean that the electric (magnetic) fields are perpendicular to the direction of propagation. The waves associated with the TE (TM) fields are sometimes called the $H$ $(E)$ modes.         
\par 
Extensive and potential applications of SPPs and  DSPs demand proper theoretical investigations together with convenient and controllable tools and techniques for coupling of EM waves and surface plasmons. Also, it has become possible to control the permittivity function of materials, and thereby enabling new approaches for the excitation of SPPs and DSPs due to the availability of non-conventional plasmonic materials such as transparent conductive materials and highly doped semiconductors \cite{park2015,west2010}.  Efficient excitation of such surface waves   have become possible with specially designed structures, e.g.,  metallic gratings \cite{ritchie1968,takayama2018,ma2018}, nanoslits \cite{shi2005} and  uniaxial crystals \cite{li2008,moradi2018}.  We mention that crystals are naturally anisotropic because their internal microstructures are  asymmetric configurations of lattice patterns. They have distinct axis directions.  In contrast to an isotropic medium, where the wave has the same speed in different directions,   when light travels in crystals, depending on their composition and structure, two waves are generated which propagate at different speeds and which exhibit two different refractive indices [$n_o$ (ordinary) and $n_e$ (extraordinary)], i.e., the waves have mutually orthogonal linear polarization. This phenomenon is known as double refraction or \textit{birefringence}. However, it is also possible to induce anisotropy in  certain media, e.g., liquid crystals by the application of an external electric field.  Note that the uniaxial crystal may not be the only choice for the existence of DSWs,   these waves also emerge in the case of a biaxial crystal \cite{walker1998} or a structurally chiral material \cite{gao2010}.    
\par 
The necessary conditions together with the parameter regimes for the existence of DSPs  and their dispersion properties at the interface of a metal and a  uniaxial crystal has been studied by Li {\it et al.} \cite{li2008} with the simple  Drude model for $\epsilon_p$ (without any absorption or damping constant). In an another work, Moradi \textit{et al.} \cite{moradi2018} studied the similar theory of DSPs   with the Drude model but in a doped InSb plasma and a uniaxial rutile (TiO$_2$) crystal at THz frequencies \cite{moradi2018}. However, the theory of DSPs at optical frequencies, especially in the short-wavelength spectra has not been advanced with the DL model to account  for the contribution of both the  intraband and higher-order interband transitions of electrons in metals.   
\par 
In this work, our aim is to consider the DL model for the complex material permittivity, which holds for a wide range of wavelength spectra, and study the dispersion properties of DSPs   at the interface of an isotropic gold metal and an anisotropic uniaxial crystal TiO$_2$.   We show that  the DL model is much more pronounced than the Drude model in the regimes of short-wavelength spectra ($\lesssim 500$ nm), and the dispersion properties of DSPs are significantly modified.  
 \section{Theoretical formulation} \label{sec-th-formuln}
\begin{figure}[ht]
\begin{center}
\includegraphics[trim=1in 3in 0 2.5in, clip, width=4.5in,height=2.7in]{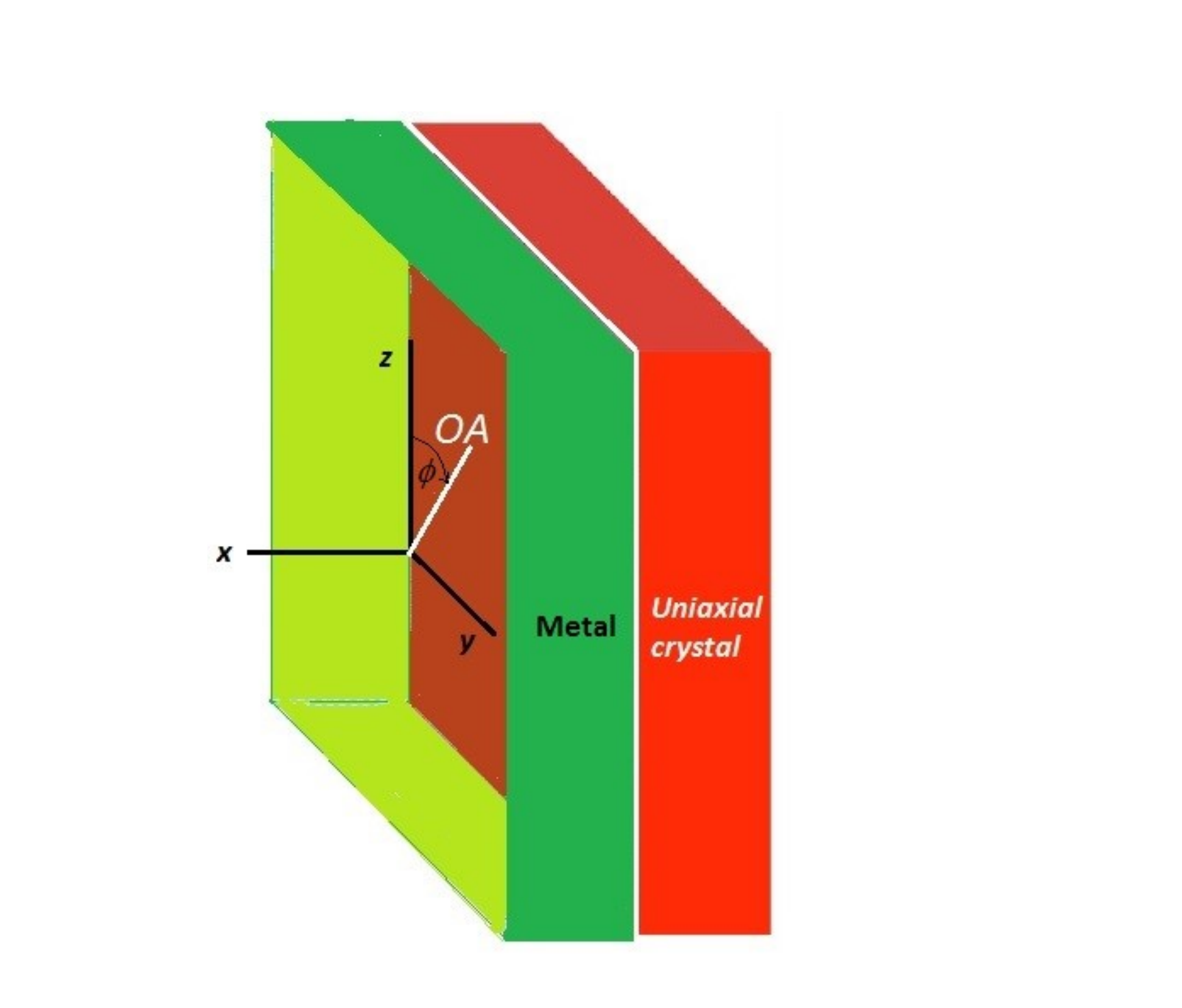}
\caption{A schematic diagram of a planar interface $(x=0)$ between  an isotropic metal   $(x>0)$ and an anisotropic uniaxial crystal $(x<0)$ is shown. Here,  $OA$ is the optical axis making an angle $\phi$ with the direction of propagation, i.e.,  the $z$-axis.}
\label{fig-diagram}
\end{center}
\end{figure}

We consider a planar interface $(x=0)$ of two media consisting of a semi-infinite isotropic  metal with permittivity $\epsilon_p$ occupying the  space $(x>0)$  and a semi-infinite anisotropic uniaxial optical crystal $(x<0)$ with the permittivity tensor $\epsilon_c$ and the optical axis $OA$ of the crystal lying in the interface plane, i.e., the $yz$-plane. The principal diagonal elements of $\epsilon_c$ are $\epsilon_o$, $\epsilon_o$ and $\epsilon_e$ which represent the  dielectric constants of the ordinary and extraordinary modes in the crystal   \cite{dyakonov1988,li2008,moradi2018}.   We also assume that the DSPs propagate    along the $z$-axis (with wave frequency $\omega$ and wave number $q$), making an angle $\phi$ with  $OA$.      A schematic diagram of the system configuration is shown in Fig. \ref{fig-diagram}. The traditional SPPs, which involve  both the plasma motion and EM waves, are known to be excited by the pure TM mode. However, the DSPs, as in Fig. \ref{fig-diagram}, may not be excited either by the pure  TE   or the  pure TM wave fields, but by both the fields.  In contrast to  two evanescent fields of usual SPPs, the DSPs, which has a polarization hybridized nature,  can have four evanescent wave fields: two TE and TM-like modes with an identical wave vector   ${\bf k}_p=(ik_p,0,q)$  and two ordinary-light (OL) and extraordinary-light (EL)-like modes with wave vectors ${\bf k}_o=(-ik_o,0,q)$ and ${\bf k}_e=(-ik_e,0,q)$. Here, $k_p,~k_o$ and $k_e$ are determined by the following dispersion laws \cite{li2008,moradi2018,dyakonov1988,carlos2018}. 
\begin{equation}
k_p^2=q^2-\epsilon_p, \label{eq-kp}
\end{equation}  
\begin{equation}
k_o^2=q^2-\epsilon_o, \label{eq-ko}
\end{equation}  
\begin{equation}
(q^2\sin^2\phi-k_e^2)/\epsilon_e+(q^2\cos^2\phi)/\epsilon_o=1. \label{eq-ke}
\end{equation}  
 The permittivity function of the metal can be described by the Drude-Lorentz model as \cite{rakic1998}
\begin{equation} 
\epsilon_p(\omega)=\epsilon_{\infty}-\frac{\Omega^2_p}{\omega(\omega+i\gamma_d)}+\sum_{j=1}^{m}\frac{f_j\omega_p^2}{\omega_j^2-\omega(\omega+i\omega\gamma_j)}.\label{eq-DL} 
\end{equation}
The combination of the first and the second terms on the right-hand side of Eq. \eqref{eq-DL}  is referred to as the \textit{intraband} part (i.e., the Drude model with free-electron effects) and the third term  as the \textit{interband} part (i.e., the Lorentz model with the effects of bound electrons). Also,  $\epsilon_\infty$ stands for the relative permittivity of the metal at high (infinite) frequency,  $\omega_p$ is the plasma oscillation frequency of the bulk metal, $\Omega_p=\sqrt{f_0}\omega_p$ is the plasma frequency associated with the intraband transitions with oscillator strength $f_0$ and $\gamma_d$ is the damping constant.  Here, $f_0$ depends on the effective mass of electrons. Furthermore,   the summation is considered to take into account the effects of multiple interband transitions of electrons in which  $m$ is the number of high-energy oscillators with the resonant frequency $\omega_j$, weighting coefficient $f_j$ and lifetime $1/\gamma_j$ for $j=1,2,...,m$. The physical interpretations  of the real and imaginary parts of the dielectric function are that while the real
part determines the degree of polarization when the material is subjected to an electric field or magnetic field, the imaginary part determines that of absorption inside the medium.
\par 
As stated before, the Drude model is the simplest description of the permittivity function of metals and semiconductors, and it  holds for the intraband transition  of  free or conduction electrons. To account the net contribution of positive ion core,  the parameter $\epsilon_\infty$ may be introduced in the Drude model \cite{aneto2017}. Though, the Drude model  has been used in many physical situations to fit with experimental results (especially in the regimes of longer infrared wavelengths, i.e.,   $\lambda>500$ nm \cite{li2017}), it has some limitations, e.g., it diverges as $\omega\rightarrow0$, it does not include the wave vector and it may not be valid for a short-wavelength spectra ($\lesssim 500$ nm) where the contribution of interband transitions become important. However, in the present work,   the very low-frequency limit, i.e.,  $\omega\rightarrow0$ is not relevant as $\omega\lesssim\omega_p$. However,  the divergence issue, if any, may be resolved  by replacing the Drude model by the Lindhard model \cite{aneto2017}.  
\par 
In order that the DSPs exist at the interface of two media, the real parts of the wave vector components $k_p$, $k_o$, $k_e$ and $q$ must positive. As in Refs. \cite{dyakonov1988,li2008,moradi2018}, the propagation angle  is considered to be in the regime $0\le\phi\le\pi/2$. Applying the appropriate boundary conditions,  namely, the tangential components of the electric and magnetic fields are continuous at the interface $x=0$,   one can obtain the following dispersion relation for DSPs \cite{dyakonov1988,li2008,moradi2018}.
\begin{equation}
(k_p+k_e)(k_p+k_o)(\epsilon_pk_o+\epsilon_ok_e)=(\epsilon_e-\epsilon_p)(\epsilon_p-\epsilon_o)k_o. \label{eq-disp-rln}
\end{equation}  
We note that the permittivity function in Eq. \eqref{eq-DL}  is complex with its real part may be positive or negative depending on the values of $\epsilon_\infty$ and other parameters. However, we consider the case in which  $\omega\lesssim\omega_p$ (where the optical properties of a medium exhibit metal-like behaviors) and $\Re \epsilon_p<0$, $\Re \epsilon_o,~\Re \epsilon_e>0$  as in Refs. \cite{dyakonov1988,li2008,moradi2018}.      The  conditions for the existence  of DSPs in absence of any collision and using the Drude model has been discussed in detail in Ref. \cite{li2008}. We, however, consider the collisional and resonance effects, and analyze the dispersion relation \eqref{eq-disp-rln} numerically in the next section \ref{sec-results}. 
\section{Results and discussion} \label{sec-results}
In this section, we study  the dispersion properties of Dyakonov surface plasmon oscillations that can propagate at the interface of an isotropic gold metal (Au) and a uniaxial crystal (TiO$_2$)  using the DL model for the  permittivity of metals. To this end, we numerically solve Eq. \eqref{eq-disp-rln} together with Eqs. \eqref{eq-kp} to \eqref{eq-ke}  using MATLAB for implicit functions, and consider the parameters that are relevant for gold metal  \cite{rakic1998} and uniaxial rutile \cite{parameter-rutile}. 
 The advantage of  considering a uniaxial crystal  is that it has a single optical axis   and two of its refractive indices are equal  $n_{xx}=n_{yy}\equiv n_o\neq n_{zz}\equiv n_e$ in contrast to the biaxial crystals $(n_{xx}\neq n_{yy},~n_{xx}\neq n_{zz})$ having two optical axes. One positive uniaxial crystal $(n_e>n_o)$ is rutile which has not only one of the highest refractive indices at visible wavelengths of any known crystal but also exhibits high dispersion and a   large birefringence. Due to these important properties, rutile is useful in polarization optics for the manufacture of certain optical elements at   visible and near infrared wavelengths. On the other hand,  any one of the noble metals, namely gold, silver, aluminium or copper can be considered as an isotropic medium due to their high optical properties, e.g., their reflection of infrared rays are almost the same.   However, gold has some  special optical properties because of its exceptionally high
chemical resistance in any type of surroundings compared to the other noble metals (e.g., silver tarnishes and forms silver sulphide while aluminium and copper oxidize in air. Any chemical change on the surface of a metal significantly affects its optical properties.   
\par 
All the freuencies including the surface wave frequency and the collisional frequency are normalized by the plasma frequency $\omega_p$, whereas the wave vector components are normalized by $\omega_p/c$, where $c$ is the speed of light in vacuum. Since $\epsilon_p$ is complex due to the damping constant $\gamma_d$, the other permittivity constants  $\epsilon_o$ and  $\epsilon_e$, as well as the wave vector components $k_p,~k_o,~k_e$ and $q$ are also  complex quantities. For the gold metal  at room temperature, we consider the plasma density as $n=6\times10^{28}$ m$^{-3}$  such that $\omega_p=13.8\times10^{15}$ s$^{-1}$. Also, we choose $\epsilon_\infty=1.2$, $f_0=0.760$ and $\gamma_d=0.0058$ and the other parameter values as given in Table \ref{table-param} and/or  in Ref. \cite{rakic1998}.   The parameters for TiO$_2$ are considered as \cite{parameter-rutile} $\epsilon_o=(2.89)^2+i (0.02)^2$ and $\epsilon_e=(3.3)^2+i(0.03)^2$ at the frequency, $f\sim0.36$ (in terms of units,  $f\sim5\times10^{15}$ s$^{-1}$ or the wavelength  $\sim2\pi c/f\approx 400$ nm.   Due to limited source of experimental data for the DL model, we consider the effects of at most five interband transitions $(m=5)$ in the gold metal.
\par
\begin{figure*}
\includegraphics[scale=0.38]{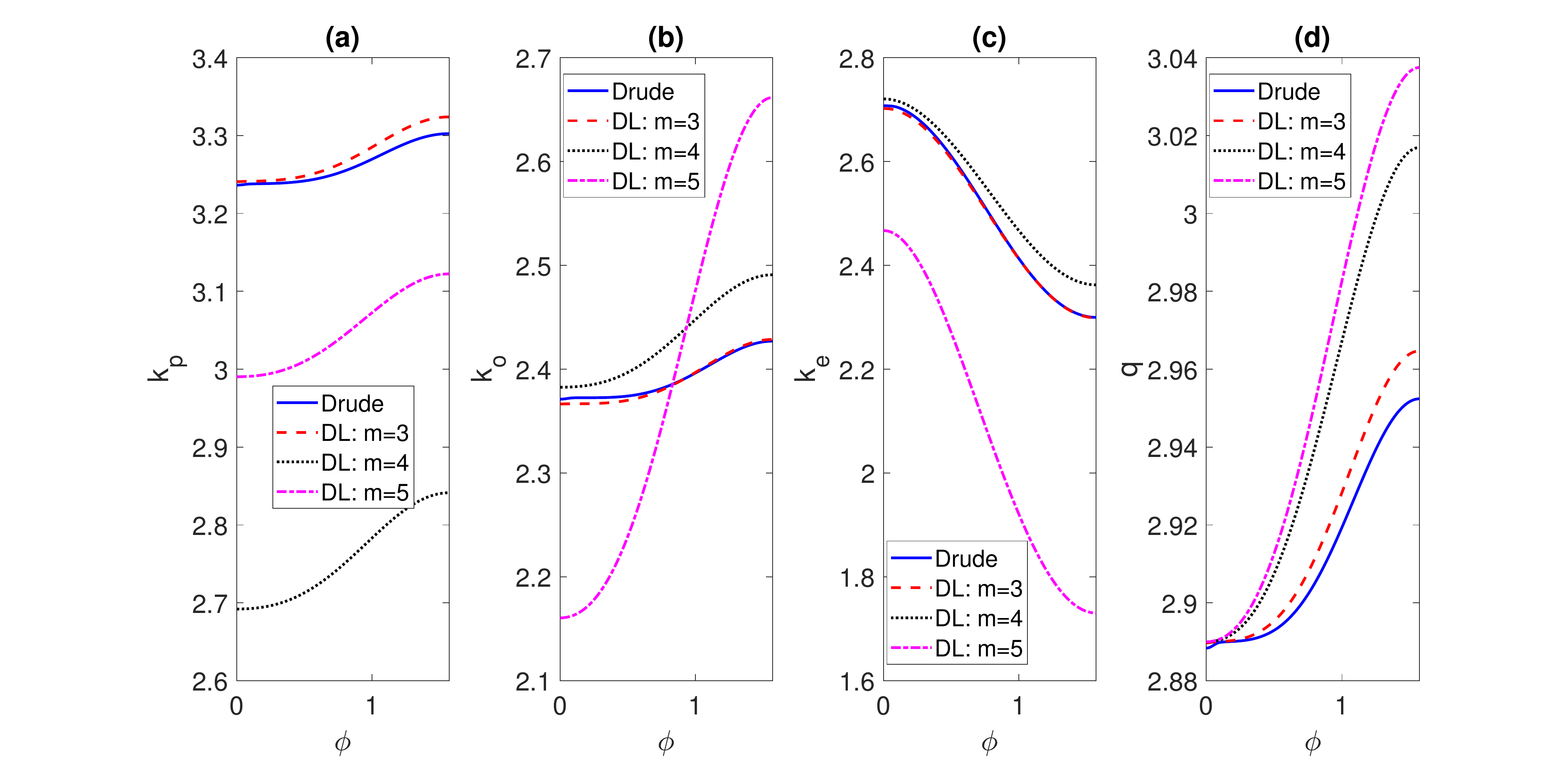}
\caption{The real parts of the wave numbers (a) $k_p$, (b) $k_o$, (c) $k_e$ and (d) $q$ are plotted against the propagation angle $\phi$ to show the effects of multiple interband transitions for a particular value of the wave frequency $\omega=0.36$ at which $\epsilon_0=(2.89)^2+i(0.02)^2$ and $\epsilon_e=(3.3)^2+i(0.03)^2$ relevant for TiO$_2$ uniaxial crystals \cite{parameter-rutile}.  Also, $\epsilon_\infty=1.2$ and other parameter values  relevant for gold metals are given in Table \ref{table-param}. The dash-dotted lines in   subplots (b) and (c) are scaled as $2k_o$ and $2k_e$ respectively.  }
\label{fig-waveno1}
\end{figure*}

Figure \ref{fig-waveno1} displays the characteristics of the real parts of the wave vector components  $k_p,~k_o,~k_e$ and $q$ against the propagation angle $\phi$ corresponding to the Drude model (solid lines) and the DL model (dashed, dotted and dash-dotted lines) for the permittivity $\epsilon_p$.   The effects of  the intraband transitions of free electrons (solid lines) and  the combined effects of both the intraband  and multiple interband transitions of electrons (dashed, dotted and dash-dotted lines) are shown.  In  the short-wavelength spectra (below $500$ nm),  we choose $\omega=0.36$ (i.e., the wavelength, $\lambda\equiv 2\pi c/\omega\sim400$ nm) at which the values of the real parts of $\epsilon_p$   for the Drude model and the DL model with $m=1,...,5$    are negative.       We find that the Drude model and the DL model with one interband transition predict almost the same  behaviors.   However, significant changes occur    when  more than three interband transitions are taken into account (see the dotted and dash-dotted lines). In all the cases, the wave number $k_e$ decays, but $k_p$, $k_o$ and $q$ increase with increasing values of $\phi$ within the interval $0\leq\phi\leq\pi/2$.     It is interesting to note that while the magnitudes of the wave number $q$ increase   with the effects of different multiple interband transitions [see subplot (d)], those of $k_p$, $k_o$ and $k_e$ may increase or  decrease [see subplots (a) to (c)]  depending on the values of $m=3$, $4$ or $5$. It means  that the DSPs are sensitive to the change of permittivity  either of  the isotropic metal or of the anisotropic crystal, and that because of the decreasing natures of  $k_e$ in the rutile substrate   and increasing natures of $k_p$ or $q$ in the metal, the DSPs may be said to be moderately localized \cite{takayama2017}. 
  Note here that these qualitative features, as shown in Fig. \ref{fig-waveno1}, remain almost the same in a wide range of  wavelength spectra $200$ to $650$ nm or the frequency range $3.8$ to $9.4\times10^{15}$ $s^{-1}$. 
  \par
\begin{figure*}[ht]
\includegraphics[scale=0.38]{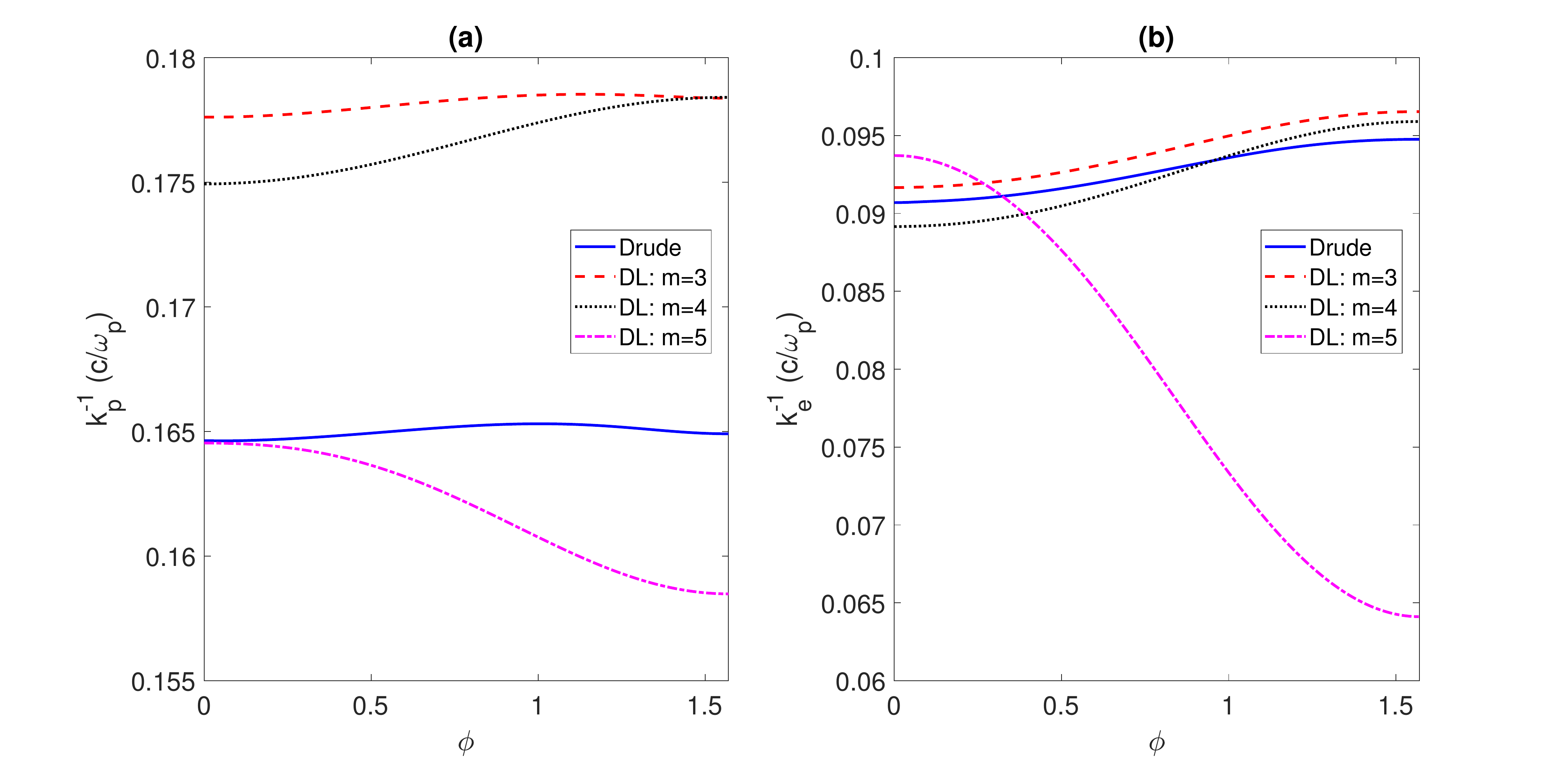}
\caption{The penetration depths  of DSPs are shown (a) in the isotropic metal (b) in the uniaxial crystal for different interband transition effects to the material permittivity. The dash-dotted lines are scaled as $k_p^{-1}/2$  and $k_e^{-1}/4$  }
\label{fig-penetration}
\end{figure*}

 
 One important characteristic of DSPs is the penetration depth or skin depth, i.e.,  $k_p^{-1}$ in the metallic gold and $k_e^{-1}$ in the rutile substrate.  Since  the penetration depth determines the coupling strength between the surface plasmons and other elements of photonic materials, its enhancement   is most desirable. From the subplots (a) and (b) of Fig. \ref{fig-penetration}, it is clear that the penetration depths in the metal and uniaxial crystal increase  with increasing values of $\phi$ except those at $m=5$ in which they decrease with increasing values of $\phi$. Such an enhancement of the penetration depth for $m=1$ to $4$ reaches maximum for perpendicular  orientation $(\phi=\pi/2)$ and minimum for parallel   orientation $(\phi=0)$ of the optical axis $OA$. So, it follows that the perpendicular  orientation is more preferable for transmission of signals in the optical wavelength. Furthermore, the penetration depth is higher at $m=3$ and $m=5$ interband transitions compared to that at $m=4$ and no interband transition. The enhancement is significant at $m=5$ (see the dash-dotted lines, scaled as $k_p^{-1}/2$ and $k_e^{-1}/4$) though at this transtition the penetration depth decreases with $\phi$ which may favor better confinement of DSPs to the interface \cite{takayama2017}. Thus, in contrast to four interband transitions,  when five or higher interband transitions come into play in metals   the parallel  orientation is more desirable than the perpendicular orientation  for transmission of signals.        
\par
\begin{figure*}
\includegraphics[scale=0.38]{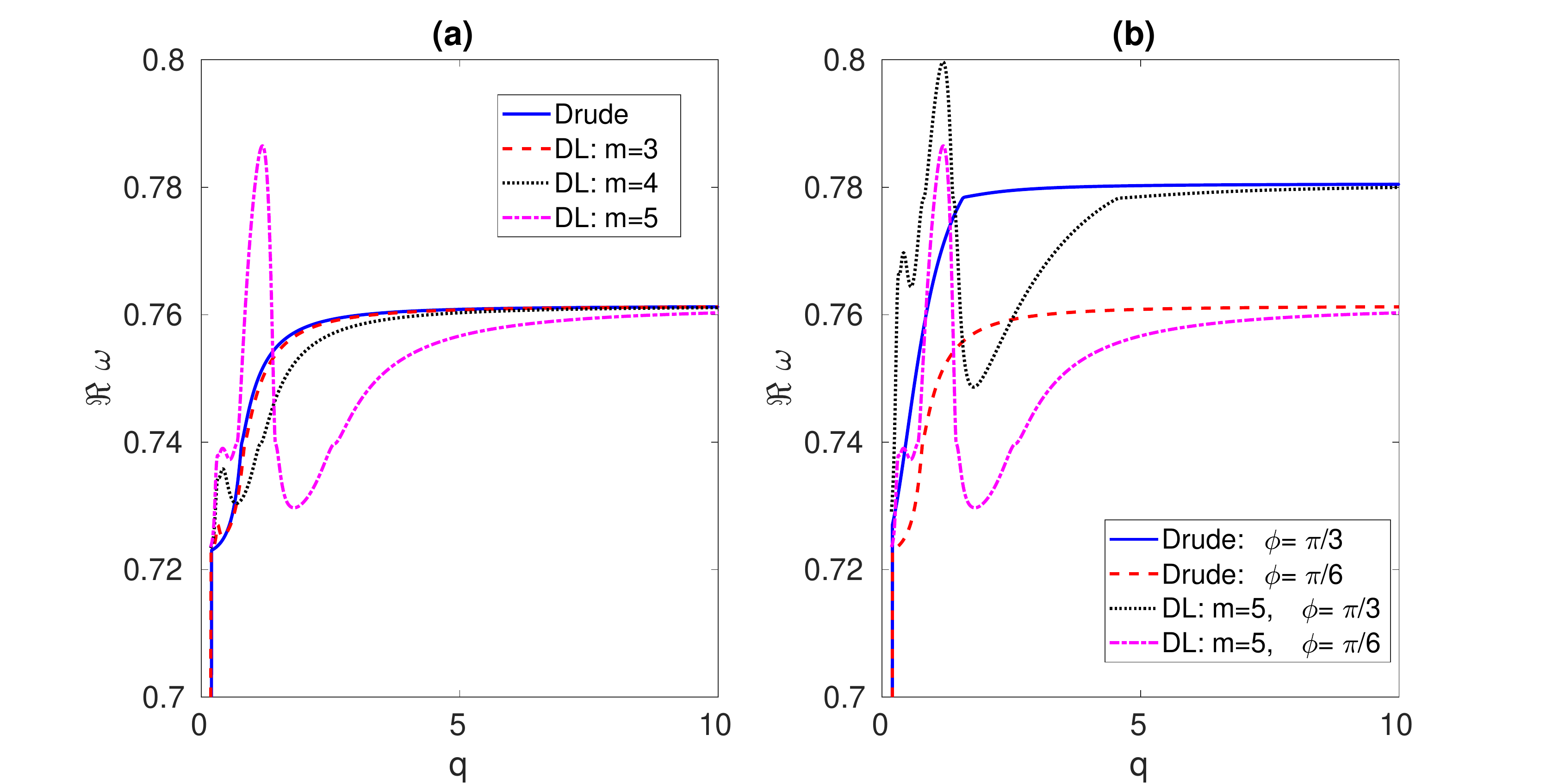}
\caption{Plots of the real part of the wave frequency  against the wave number $q$ of DSPs  to show the effects of (a) the multiple interband  transitions for a fixed $\phi=\pi/6$ and (b) the angle of propagation. We choose $\epsilon_0=(2.89)^2+i(0.02)^2$ and $\epsilon_e=(3.3)^2+i(0.03)^2$ relevant for TiO$_2$ uniaxial crystals at a frequency $\sim3\times10^{15}$ s$^{-1}$ \cite{parameter-rutile}.  Also, we consider $\epsilon_\infty=10.0$ and other parameter values  relevant for gold metals as given in Table \ref{table-param}. }
\label{fig-frequency}
\end{figure*}

 
 The dispersion curves, i.e., the plots of the wave frequency (real part)  against the wave number $q$ of DSPs are shown in Fig.  \ref{fig-frequency}  in two different cases: (a)  the effect of multiple interband transitions and (b)  the effect of the angle of propagation or the orientation of the optical axis.   It is to be noted that different values of $\epsilon_\infty$   gives rise different resonance frequencies, i.e., a modification of $\epsilon_\infty$ $(\sim10)$   tunes the resonance in a wide range of frequency. This allows to choose a suitable value of $\epsilon_\infty$ in order to match the theoretical results with  experimental ones.  For example, some authors considered $\epsilon_\infty=11.5$ for $\lambda>516$ nm and  $\epsilon_\infty=10$ for $400\lesssim\lambda\lesssim500$ nm  etc. \cite{alabastri2013}.  
 Here, we have considered   $\epsilon_\infty$ $(\sim10)$  in order to clearly exhibit   the resonance effects, i.e., multiple peaks within a short range of values of $q$.   However, the  qualitative behaviors of the dispersion curves with some lower values of $\epsilon_\infty$ than $10$ remain almost the same.   From the subplot (a), it is found that when no interband transition is considered (the solid line), the behavior of the dispersion curve remains the same as in Refs. \cite{li2008,moradi2018}, i.e., the wave frequency approaches a constant value after it starts increasing  within a  short-range of values of  $q$.  However,  when multiple interband transitions are considered together with the intraband transition, the DL model gives a significant modification of the dispersion curves, especially for $(m>3)$.   In contrast to the Drude model and previous investigations \cite{li2008,moradi2018}, the DSPs clearly display dispersion as well as resonant behaviors  within a short-range of values of $q$ (see the dotted and dash-dotted lines). Physically, these occur due to the contribution from interband transitions of bound electrons in the metal gold to the permittivity function $\epsilon_p$ with different oscillation frequencies $\omega_j$ of the resonant modes. 
From the subplot (a), it is also clear that the Drude model is no longer applicable when the effects of  more than three interband transitions come into the picture in the short-wavelength spectra:  $\lambda\lesssim500$ nm. Furthermore, the wave frequency decreases with the   effects of multiple interband transitions $(m>2)$ in the DL model except in the  regime of small $q$, i.e., $q\lesssim1$. 
\par 
The direction of propagation $\phi$ with the optical axis also plays an important role in the characteristics of DSPs as depicted in Fig. \ref{fig-frequency} (b). It is seen that  the wave frequency in both the cases (with the Drude  and the DL models) is significantly reduced with a  reduction of the angle of propagation (see the solid and dashed lines for the Drude model, and dotted and dash-dotted lines for the DL model). So, by reducing or increasing the angle of propagation from an initial value  one can have a wide range of  wavelength spectra for DSPs including the regime with $\lambda\lesssim500$ nm. In this way, the wave frequency of DSPs can be tuned with changing the orientation of the optical axis.   

\begin{table}[ht]
\caption{The parameter values relevant for the gold metal (Au) as in Ref. \cite{rakic1998} for the Drude-Lorentz model with $\gamma_d=0.053$ ev [$0.0805\times10^{15}$ s$^{-1}$] and $\omega_p=9.1$ ev [$13.8\times10^{15}$ s$^{-1}$] for which $n=6\times10^{28}$ m$^{-3}$.  All the frequencies are normalized by $\omega_p$.}  
\centering                     
\begin{tabular}{|c|c|c|c|c|}       
\hline                   
$f_j$ & $\gamma_j$ & $\gamma_j$  & $\omega_j$  & $\omega_j$  \\   
& (ev)&$(10^{15}$ s$^{-1})$& (ev) &$(10^{15}$ s$^{-1})$\\ \hline
$0.024$ & $0.241$ & $0.3661$ & $0.415$ & $0.6304$\\ 
$0.010$ & $0.345$ & $0.5241$ & $0.830$ & $1.2609$\\  
$0.071$ &  $0.870$ & $1.3216$ & $2.969$ & $4.5102$ \\  
$0.601$ & $2.494$ & $3.7886$ & $4.304$ & $6.5382$\\  
$4.384$ &  $2.214$ & $3.3633$ & $13.32$ & $20.2344$\\
[1ex]   
\hline 

\end{tabular}
\label{table-param}    
\end{table}
 \par
\begin{figure*}
\includegraphics[scale=0.38]{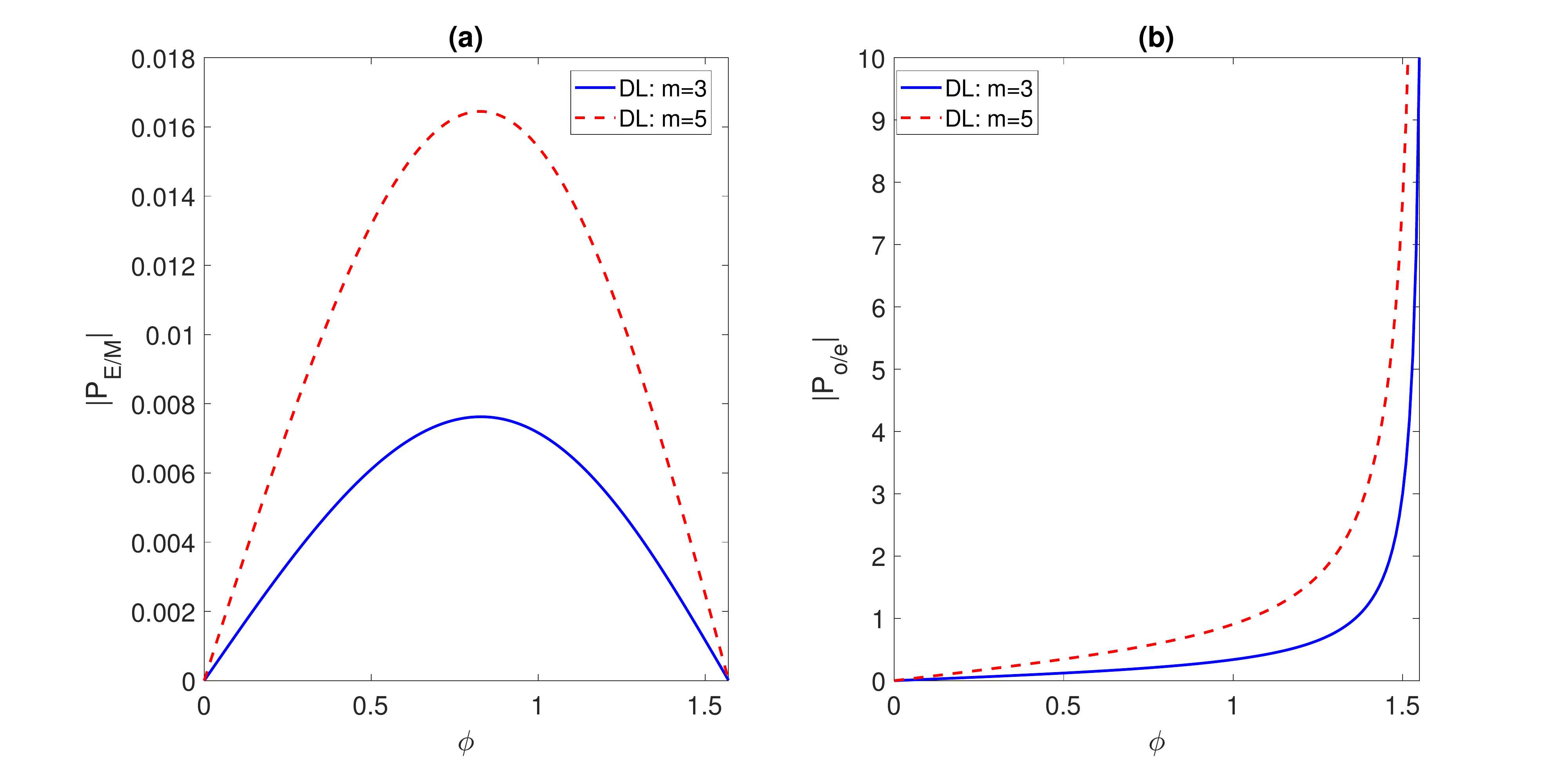}
\caption{The absolute values of the hybridization factors (a) $P_{E/M}$ and (b) $P_{o/e}$ are plotted against the propagation angle $\phi$. The effects of the interband transitions with $m>3$ are seen to be significant.    }
\label{fig-pemoe}
\end{figure*}

 
We have mentioned that the DSPs are hybridized due to polarization of both the TE and TM modes. In order to distinguish the polarization characteristics, we define
two factors $P_{E/M}$ and $P_{o/e}$, respectively, as the amplitude ratios between the  TE and TM modes in the isotropic metal   and between the ordinary and extraordinary modes in the uniaxial crystal as \cite{moradi2018}
\begin{equation}
P_{E/M}=\frac{q^2-k_ek_o-\epsilon_o}{i\epsilon_o(k_e+k_p)\tan\phi+ik_e(\epsilon_o-q^2-k_ok_p)\cot\phi}, \label{eq-pem}
\end{equation}
\begin{equation}
P_{o/e}=\frac{i(k_p+k_e)\tan\phi}{\epsilon_o-q^2-k_ok_p}.\label{eq-poe}
\end{equation}
The absolute values of $P_{E/M}$  and $P_{o/e}$ are plotted against $\phi$ as shown in Fig. \ref{fig-pemoe}. It is found that for these ratios,   the Drude model and the DL model with three interband transitions $(m=3)$ predict almost the same results, however, it is significantly modified with $m>3$ [see the dashed lines in subplots (a) and (b)]. The values of both the ratios $|P_{E/M}|$ and $|P_{o/e}|$  increase with the effects of higher-order   interband transitions of electrons. However, the ratio $|P_{E/M}|$ reaches its maxium at an intermediate value of $\phi$ and then decreases   to have a cut-off  at $\phi=\pi/2$. Such cut-offs may be different for different values of either $\epsilon_p$ or both   $\epsilon_o$ and $\epsilon_e$. Although, the DSPs have been known to be TM-dominant \cite{moradi2018},    such an increase of $|P_{E/M}|$ in the present analysis indicates that  the contribution  of  the TE-polarization may no longer  be negligible, especially at higher-order interband transitions of electrons.  On the other hand, the effect  of the interband transitions on the ratio $|P_{o/e}|$ in the uniaxial crystal is also found to be significant and that  its values  become  higher and higher with increasing values of $\phi$. 
 \section{Conclusion}\label{sec-conclu}
We have studied  the dispersion properties of Dyakonov surface plasmons (DSPs) propagating along the interface of an isotropic metallic plasma [gold (Au)]  and an anisotropic  uniaxial crystal [Rutile (TiO$_2$)] using the Drude-Lorentz (DL) model for the dielectric permittivity $\epsilon_p$ of metals. The DL model is considered to take into account the effects of both the \textit{intraband} transition of free electrons and \textit{interband} transitions of   bound electrons in metals. The higher-order interband transitions of bound electrons are very common especially in practical metals and these  occur due to the excitation by the high-energy photons. In the short-wavelength spectra ($\lambda\lesssim500$ nm), since the interband transitions occur in the gold metal, the DL model is more appropriate than the Drude model for the description of $\epsilon_p$.  It is found that the contributions of multiple interband transitions, as well as, the orientation $(\phi)$ of the optical axis  significantly modify the wave vector components $k_p$ and $q$  of the TE/TM modes in the isotropic metal  and $k_o$ and $k_e$ of the ordinary and extraordinary modes in the anisotropic crystal. The modifications are noticeable when the contribution of three or more interband transitions come into the picture. The  values of the wave numbers decrease or increase depending on the values of $\phi$ and the order of interband transitions $m$ in the permittivity $\epsilon_p$.  We have also shown that   the penetration depths of DSPs in the metal and uniaxial crystal increase with increasing values of $\phi$ in the whole interval $0\leq\phi\leq\pi/2$ and when the effects of multiple interband transitions $(m<5)$ are considered. However, the exception occurs at $m=5$ in which case the penetration depth decreases with increasing values of $\phi$.  The enhancement of the penetration depth becomes significantly higher at $m=3$ and $m=5$ compared to those at $m=1$ to $4$ and with no interband transition.     So, an appropriate choice of the orientation of the optical axis and the multiple   interband transitions of electrons may be required for the (i) better confinement of DSPs to the interface having lower penetration depth and (ii) transmission of optical signals with higher penetration depth of DSPs.     
\par 
A numerical solution of the dispersion equation  also reveals that in contrast to the Drude model or DL model with upto three interband transitions, the real part of the wave frequency of DSPs exhibits strong dispersion and resonance with multiple peaks at higher-order $(m>3)$ interband transitions. We have also calculated the absolute amplitude ratios between the  TE and TM modes in the isotropic metal $(|P_{E/M}|)$, as well as between the ordinary and extraordinary modes in the uniaxial crystal $(|P_{o/e}|)$. It is  shown that in contrast to the previous investigation \cite{moradi2018} with the Drude model in semiconductor plasmas where the contribution of the TE-mode was reported to be negligible in the excitation of DSPs,    the contribution of the TE mode to DSPs may not  be  negligible  when  more than three interband transitions   are taken into account.     
\par 
To conclude, the DSPs can be used as a sensor and switching system  due to their high sensivity to the relative values of the permittivities of the two media. Also, since the wave frequency can be tuned with changing the orientation of the optical axis, the DSPs may be a good candidate for the  transportation of  directional EM signals. The theoretical results should be useful to design new experiments for the excitation of surface EM waves that can propagate along the interface of an isotropic gold metal and an anisotropic uniaxial crystal (rutile) at optical frequencies. Finally, it has been found in Ref. \cite{alabastri2013} that any change of thermodynamic temperature can have strong influence on the field enhancement and absorption characteristics of plasmonic devices. In the light of this research, the present work could be advanced with a more general model by considering the temperature dependency of the plasma frequency $\Omega_p$ and the damping coefficient $\gamma_d$ in the Drude model, i.e.,  the Drude-Lorentz-Temperature (DLT) model  for the material permittivity  $\epsilon_p$. However, this study is left for future work. 
\section*{Acknowledgments} {This work was supported by Science and Engineering Research Board (SERB), Govt. of India with  Sanction  order no. CRG/2018/004475   dated 26 March 2019.  }

\end{document}